\documentclass[a4paper,fleqn,usenatbib]{mnras}

\usepackage{newtxtext,newtxmath}
\usepackage{lineno}
\usepackage[T1]{fontenc}
\usepackage{ae,aecompl}

\usepackage{graphicx}	% Including figure files
\usepackage{natbib, xcolor}
\newcommand{\tvcol}{TV~Col}
\newcommand{\eiuma}{EI~UMa}

\newcommand{\tess}{{\sl TESS\/}}
\newcommand{\swift}{{\sl Swift\/}}
\newcommand{\xmm}{{\sl XMM-Newton\/}}

\newcommand{\asca}{{\sl ASCA\/}}

\newcommand{\nustar}{{\sl NuSTAR\/}}

\newcommand{\gaia}{{\sl Gaia\/}}

\newcommand{\eps}{erg\,s$^{-1}$}

\newcommand{\pspin}{P$_{\rm s}$}
\newcommand{\porb}{P$_{\rm o}$}
\newcommand{\pbeat}{P$_{\rm b}$}
\newcommand{\gabs}{M$_{\rm G}$}
\newcommand{\qmag}{M$_{\rm V,q}$}
\newcommand{\omag}{M$_{\rm V,o}$}

\title[Absolute Magnitudes of Intermediate Polars]{The Orbital Period vs. Absolute Magnitude Relationship of Intermediate Polars: Implications for Low States and Outbursts}

\author[K. Mukai \& M.L. Pretorius]{Koji Mukai,$^{1,2}$\thanks{E-mail: Koji.Mukai@nasa.gov (KM)}
Magaretha L. Pretorius$^3$
\\
$^{1}$CRESST II and X-ray Astrophysics Laboratory, NASA/GSFC, Greenbelt, MD 20771, USA\\
$^{2}$Department of Physics, University of Maryland Baltimore County, 1000 Hilltop Circle, Baltimore MD 21250, USA\\
$^{3}$South African Astronomical Observatory, PO Box 9, Observatory 7935, South Africa\\
}

\date{Accepted XXX. Received YYY; in original form ZZZ}

\pubyear{2022}

\begin{document}
\label{firstpage}
\pagerange{\pageref{firstpage}--\pageref{lastpage}}
\maketitle

\begin{abstract}
Recent advances in time-domain astronomy have led to fresh observational
insights into intermediate polars, a subtype of magnetic cataclysmic
variables generally accreting via a partial accretion disc.
These new discoveries include detections of superhumps, low states,
and outbursts. However, these studies have largely relied on relative
photometry. Here we tabulate the absolute G magnitudes of confirmed
intermediate polars, plot them against their orbital periods, and
compare the results to similar studies of dwarf novae during quiescence
and in outburst. This exercise suggests the presence of two distinct
luminosity classes of intermediate polars, with practical and physical
implications for the studies of low states and outbursts. In particular,
we point out that two of the optically luminous systems showing short
outbursts are also seen to exhibit superhumps, suggesting
that they may be caused by the same underlying mechanism.
\end{abstract}


%%%%%%%%%%%%%%%%%%%%%%%%%%%%%%%%%%%%%%%%%%%%%%%%%%

%%%%%%%%%%%%%%%%% BODY OF PAPER %%%%%%%%%%%%%%%%%%

\section{Introduction}
\label{intro}

Cataclysmic variables (CVs) are semi-detached close binary systems
in which a Roche lobe-filling late type star (the secondary) transfers
matter onto a white dwarf (the primary).  CVs are known to vary in the
optical on a wide range of timescales, from seconds, hours, days to years
\citep{CoelBook}, making them excellent targets for time-domain surveys.

This includes observations with the Transiting Exoplanet Survey Satellite
(\tess; \citealt{TESS}). For the purpose of exoplanet search, \tess\ is designed to provide high cadence, high
precision photometry almost continuously for a
period of $\sim$28 days or integer multiples thereof, covering almost the
entire sky over the course of the mission.
This also makes it an excellent tool for the studies of CVs. Specifically,
\tess\ data enable searches for rare photometric events and also enable
significantly improved studies of multiply-periodic systems, particularly
when the frequencies are closely spaced.  The latter is the case for the
intermediate polars (IPs, also called the DQ~Her-type systems; \citealt{IPs}),
in which the white dwarf magnetic field is sufficient (of order 10 MG) to
disrupt the inner accretion disc, but not strong enough to synchronize
the white dwarf spin to the orbit. The accretion flow follows the magnetic
field lines to the magnetic pole regions in a form often called the accretion
curtains. At the footpoints of the curtains, strong shocks form and produce
X-rays \citep{Mukai2017}. The X-rays are reprocessed by the accretion
curtains, the partial disc, and the secondary surface, resulting in optical
light curves that are modulated on the spin (\pspin), the orbital (\porb),
and/or the beat (\pbeat=[\pspin$^{-1}-$\porb$^{-1}$]$^{-1}$) periods.
In addition, the dense, nearly continuous coverage with \tess\ is also
advantageous in the studies of superhumps in non-magnetic CVs
\citep{Bruch2022}, seen originally in SU~UMa type dwarf novae; we will
discuss this in the context of IPs in subsection\,\ref{IPSH}.

Most other time domain surveys do not have the dense coverage necessary
to track the orbital, spin, and beat periods of IPs. However, surveys
such as All-Sky Automated Survey for Supernovae (ASAS-SN;
\citealt{Shappee2014,Kochanek2017}) and the observations by the
amateur astronomers, as tabulated by the American Association
of Variable Star Observers (AAVSO\footnote{https://www.aavso.org/})
have longer baselines and allow searches for outbursts and low states.

Between \tess, ASAS-SN, and the AAVSO database, these time-domain
observations have resulted in several recent publications on outbursts (see, e.g.,
\citealt{Scaringi2022Nature,Hameury2022}) and low states (e.g.,
\citealt{Covington2022}) of IPs. However, these studies
have concentrated on relative photometry. Here we collect absolute
magnitudes of IPs based on \gaia\ data and present them as a function
of \porb. This exercise allows us to place IPs on the same plane as
other CVs with accretion discs, and hence provides the necessary context
in which to interpret the low states and outbursts.

\section{Absolute Magnitudes vs. Orbital Periods of IPs}

For this compilation, we started with the list of 71 objects confirmed
as IPs as of late
2021\footnote{https://asd.gsfc.nasa.gov/Koji.Mukai/iphome/catalog/alpha.html}.
Of these, 67 have positive parallax measurements in \gaia\ early
Data Release 3 (EDR3; \citealt{EDR3}). as well as \gaia\ G
magnitudes (the G band covers 330--1050 nm; \citealt{GaiaMission}).
G magnitudes reported in the \gaia\ EDR3 catalog are the weighted means
of calibrated measurements taken between 2014 July 25 and 2017 May 28
\citep{Riello2021}. For three IPs (CXOGBS~J174954.5$-$294335,
IGR~J18151$-$1052, and V1674~Her), negative parallax values are reported,
while there is no \gaia\ counterpart to XY~Ari, the IP behind the molecular
cloud, MBM~12, so these systems are not included in this analysis.  For
the rest, we used the geometric distances of \cite{GaiaEDR3}, and hence
calculated their absolute G magnitudes \gabs. The orbital period and
\gabs\ of these 67 IPs, along with several other properties explained
below, are listed in Table\,\ref{tab:ipabsmag} in order of \gabs.

\begin{table*}
%tablewidth{0pt}
%\setlength{\tabcolsep}{0.025in}
\centering
\caption{Quiescent Absolute Magnitudes of Intermediate Polars}
\begin{tabular}{lrrrrrlcc}
\hline
Object & \porb\ (h) & \gabs & DBI$^1$ & $L_{\rm HX}$ & $L_{\rm SX}$ & SX Ref & Outburst & Low State \\
\hline
1RXS J180431.1$-$273932 & 4.99 & 2.9: & 1.392 &  & 4.4$\times$10$^{34}$ & 4XMM-DR11 & N & N \\
V667 Pup & 5.61 & 3.0 & 1.326 & 1.0$\times$10$^{34}$ & 5.1$\times$10$^{33}$ & 2SXPS & N & N \\
IGR J18173$-$2509 & 1.53 & 3.5: & 1.440 & 3.9$\times$10$^{34}$ & 2.9$\times$10$^{34}$ & 4XMM-DR11 & N & N \\
V4743 Sgr & 6.72 & 3.6: & 1.084 &  & 4.8$\times$10$^{33}$ & 4XMM-DR11 & N & N \\
NY Lup & 9.87 & 3.8 & 0.784 & 1.8$\times$10$^{34}$ & 7.9$\times$10$^{33}$ & 4XMM-DR11 & Y & N \\
V1223 Sgr & 3.37 & 4.3 & 1.123 & 4.9$\times$10$^{33}$ & 3.4$\times$10$^{33}$ & 4XMM-DR11 & Y & Y \\
1RXS J213344.1+510725 & 7.14 & 4.3 & 0.856 & 1.3$\times$10$^{34}$ & 6.1$\times$10$^{33}$ & 4XMM-DR11 & N & Y \\
V2731 Oph & 15.42 & 4.4 & & 3.1$\times$10$^{34}$ & 1.1$\times$10$^{34}$ & 4XMM-DR11 & N & N \\
GK Per & 47.92 & 4.4 & & 1.3$\times$10$^{33}$ & 7.1$\times$10$^{32}$ & ASCA & Y & N \\
V418 Gem & 4.37 & 4.5: & 1.002 & 8.0$\times$10$^{33}$ & 6.2$\times$10$^{33}$ & 4XMM-DR11 & N & N \\
IGR J15094$-$6649 & 5.88 & 4.5 & 0.894 & 3.4$\times$10$^{33}$ & 2.4$\times$10$^{33}$ & 4XMM-DR11 & N & N \\
V647 Aur & 3.47 & 4.5 & 1.064 & 6.0$\times$10$^{33}$ & 3.7$\times$10$^{33}$ & 4XMM-DR11 & N & N \\
EI UMa & 6.43 & 4.6 & 0.825 & 4.3$\times$10$^{33}$ & 5.4$\times$10$^{33}$ & 4XMM-DR11 & Y & N \\
Swift J2006.4+3645 & 17.28 & 4.6: & & 3.3$\times$10$^{34}$ & 9.5$\times$10$^{33}$ & 2SXPS & N & N \\
HZ Pup & 5.09 & 4.6: & 0.924 &  & 2.2$\times$10$^{33}$ & 4XMM-DR11 & N & N \\
PQ Gem & 5.19 & 4.8 & 0.863 & 2.1$\times$10$^{33}$ & 1.8$\times$10$^{33}$ & 4XMM-DR11 & N & N \\
1RXS J230645.0+550816 & 3.26 & 4.8: & 0.999 &  & 3.6$\times$10$^{33}$ & 4XMM-DR11 & N & N \\
V405 Aur & 4.14 & 4.8 & 0.938 & 1.7$\times$10$^{33}$ & 1.2$\times$10$^{33}$ & 2SXPS & N & N \\
V2400 Oph & 3.43 & 4.9 & 0.961 & 2.8$\times$10$^{33}$ & 2.9$\times$10$^{33}$ & 4XMM-DR11 & N & N \\
AO Psc & 3.59 & 4.9 & 0.950 & 7.9$\times$10$^{32}$ & 1.2$\times$10$^{33}$ & 4XMM-DR11 & N & Y \\
MU Cam & 4.72 & 5.0 & 0.843 & 1.7$\times$10$^{33}$ & 1.1$\times$10$^{33}$ & 4XMM-DR11 & N & N \\
V515 And & 2.73 & 5.0 & 0.983 & 2.0$\times$10$^{33}$ & 2.7$\times$10$^{33}$ & 4XMM-DR11 & N & Y \\
V709 Cas & 5.33 & 5.1 & 0.770 & 4.8$\times$10$^{33}$ & 2.5$\times$10$^{33}$ & 4XMM-DR11 & N & N \\
PBC J0927.8$-$6945 & 4.79 & 5.2 & 0.784 & 1.5$\times$10$^{33}$ & 9.7$\times$10$^{32}$ & 4XMM-DR11 & N & N \\
BG CMi & 3.23 & 5.2 & 0.896 & 2.2$\times$10$^{33}$ & 2.2$\times$10$^{33}$ & 2SXPS & N & N \\
V2306 Cyg & 4.73 & 5.2 & 0.788 & 2.6$\times$10$^{33}$ & 2.8$\times$10$^{33}$ & 2SXPS & N & N \\
FO Aqr & 4.85 & 5.3 & 0.753 & 1.8$\times$10$^{33}$ & 9.8$\times$10$^{32}$ & 4XMM-DR11 & Y & Y \\
V2069 Cyg & 7.48 & 5.3 & 0.543 & 2.9$\times$10$^{33}$ & 1.5$\times$10$^{33}$ & 4XMM-DR11 & N & N \\
TV Col & 5.49 & 5.5 & 0.649 & 1.8$\times$10$^{33}$ & 1.8$\times$10$^{33}$ & 2SXPS & Y & N \\
IGR J16547$-$1916 & 3.72 & 5.5 & 0.783 & 2.9$\times$10$^{33}$ & 2.2$\times$10$^{33}$ & 2SXPS & N & N \\
V349 Aqr & 3.23 & 5.5: & 0.818 & & 1.1$\times$10$^{33}$ & 4XMM-DR11 & N & N \\
UU Col & 3.45 & 5.6: & 0.776 & & 2.2$\times$10$^{33}$ & 4XMM-DR11 & N & N \\
IGR J17014$-$4306 & 12.82 & 5.6 & & 1.3$\times$10$^{33}$ & 1.1$\times$10$^{33}$ & 4XMM-DR11 & Y & N \\
Swift J2138.8+5544 & 4.43 & 5.6: & 0.704 & & 4.1$\times$10$^{33}$ & 2SXPS & N & N \\
V1033 Cas & 4.03 & 5.6 & 0.733 & 4.1$\times$10$^{33}$ & 2.3$\times$10$^{33}$ & 4XMM-DR11 & N & N \\
V1062 Tau & 9.98 & 5.7 & 0.199 & 4.5$\times$10$^{33}$ & 4.6$\times$10$^{33}$ & 2SXPS & Y & Y \\
IGR J16500$-$3307 & 3.62 & 5.7 & 0.738 & 3.3$\times$10$^{33}$ & 1.7$\times$10$^{33}$ & 4XMM-DR11 & N & N \\
WX Pyx & $\sim$5.3 & 5.7 & & & 1.1$\times$10$^{33}$ & 4XMM-DR11 & N & N \\
IGR J08390$-$4833 & $\sim$8 & 5.8 & & 5.3$\times$10$^{33}$ & 2.7$\times$10$^{33}$ & 2SXPS & N & N \\
TX Col & 5.691 & 5.8 & 0.551 & 1.2$\times$10$^{33}$ & 1.8$\times$10$^{33}$ & 2SXPS & N & N \\
IGR J18308$-$1231 & 5.37 & 5.9: & 0.549 & 1.0$\times$10$^{34}$ & 4.6$\times$10$^{33}$ & 4XMM-DR11 & N & N \\
Swift J0717.8$-$2156 & 5.52 & 6.0: & 0.510 & 5.1$\times$10$^{33}$ & 3.3$\times$10$^{33}$ & 4XMM-DR11 & N & N \\
V1323 Her & 4.40 & 6.1 & 0.572 & & 8.0$\times$10$^{32}$ & 4XMM-DR11 & N & Y \\
AE Aqr & 9.88 & 6.1 & 0.088 & & 1.5$\times$10$^{31}$ & 4XMM-DR11 & N & N \\
DQ Her & 4.65 & 6.1 & 0.553 & & 3.4$\times$10$^{30}$ & 4XMM-DR11 & N & N \\
IGR J04571+4527 & 6.2 or 4.8 & 6.2 & & 5.1$\times$10$^{33}$ & 3.0$\times$10$^{33}$ & 4XMM-DR11 & N & N \\
IGR J17195$-$4100 & 4.01 & 6.2 & 0.576 & 1.8$\times$10$^{33}$ & 1.9$\times$10$^{33}$ & 4XMM-DR11 & N & N \\
Swift J183920.1$-$045350 & $\sim$5.6 & 6.6: & & & 1.4$\times$10$^{33}$ & 4XMM-DR11 & N & N \\
RX J2015.6+3711 & 12.76 & 6.7 & & & 1.0$\times$10$^{33}$ & 4XMM-DR11 & N & N \\
V3037 Oph & 5.72 & 7.2: & 0.163 & & 1.8$\times$10$^{33}$ & 4XMM-DR11 & N & N \\
AX J1740.1$-$2847 & 2.1? & 7.5: & & & 2.3$\times$10$^{33}$ & 4XMM-DR11 & N & N \\
LAMOST J024048.51+195226.9 & 7.34 & 7.9 & $-$0.186 & & & & N & N \\
V1460 Her & 4.99 & 7.9 & 0.038 & & 1.6$\times$10$^{30}$ & 2SXPS & N & N \\
IGR J19267+1325 & 3.45 & 8.3 & 0.068 & & 1.1$\times$10$^{33}$ & 2SXPS & N & N \\
AX J1832.3$-$0840 & & 8.3: & & 5.0$\times$10$^{33}$ & 2.2$\times$10$^{33}$ & 4XMM-DR11 & N & N \\
DO Dra & 3.97 & 8.8 & $-$0.111 & 8.2$\times$10$^{31}$ & 2.4$\times$10$^{32}$ & 2SXPS & Y & Y \\
HT Cam & 1.43 & 8.8 & 0.110 & & 8.7$\times$10$^{31}$ & 4XMM-DR11 & Y & N \\
DW Cnc & 1.44 & 9.0 & 0.060 & & 8.9$\times$10$^{31}$ & 4XMM-DR11 & Y & Y \\
1RXS J211336.1+542226 & 4.17 & 9.1 & $-$0.209 & 3.4$\times$10$^{32}$ & 4.5$\times$10$^{32}$ & 4XMM-DR11 & N & Y \\
EX Hya & 1.64 & 9.4 & $-$0.059 & 1.0$\times$10$^{31}$ & 6.5$\times$10$^{31}$ & 4XMM-DR11 & Y & N \\
PBC J1841.1+0138 & 5.33 & 9.7: & $-$0.484 & 4.3$\times$10$^{32}$ & & & N & N \\
V598 Peg & 1.39 & 9.8 & $-$0,137 & & 4.9$\times$10$^{31}$ & 4XMM-DR11 & N & N \\
CTCV J2056$-$3014 & 1.76 & 9.9 & $-$0.196 & & 1.8$\times$10$^{31}$ & 4XMM-DR11 & Y & N \\
CC Scl & 1.41 & 10.3 & $-$0.265 & & 7.9$\times$10$^{31}$ & XMM Slew & Y & N \\
AX J1853.3$-$0128 & 1.45 & 10.7 & $-$0.370 & & 3.7$\times$10$^{31}$ & 4XMM-DR11 & N & N \\
V1025 Cen & 1.41 & 10.7 & $-$0.367 & 4.2$\times$10$^{31}$ & 1.1$\times$10$^{32}$ & 4XMM-DR11 & Y & Y \\
V455 And & 1.35 & 11.6 & $-$0.588 & & 5.5$\times$10$^{28}$ & 4XMM-DR11 & Y & N \\
\hline
\multicolumn{9}{l}{$^1$Disc Brightness Index. See text for definition.}
\end{tabular}
\label{tab:ipabsmag}
\end{table*}

We argue below that an accuracy of $\pm$0.3 mag in \gabs\ is
desirable to achieve our main objective. In comparison, a distance error of
$\pm$4.7\% results in an error in \gabs\ of $\pm$0.1 mag. Reported errors are
smaller than this limit for 36 of the 67 IPs in Table\,\ref{tab:ipabsmag}.
An additional 14 IPs have distance accurate to $<$15\%, resulting in
\gabs\ error of 0.3 mag. The remaining 17 IPs have have larger errors
on the distance, and hence \gabs: these are indicated with a colon added
to the \gabs\ values. Since these IPs are mostly at distances exceeding
$\sim$2 kpc, they may also suffer significant interstellar reddening.

We also investigated the potential effects of long-term
variability by inspecting the data collected by the American
Association of Variable Star Observers\footnote{Specifically, we used
their online Light Curve Generator (https://www.aavso.org/LCGv2/).}
between 2014 July 25 (JD=2456863.5) and 2017 May 28 (2457902.5).
Most nearby IPs (e.g., 30 out of the 36 systems with distance errors
$<$4.7\%) are well covered by AAVSO observers, and in the majority of
the case, there is a slight, likely systematic, offset in that G magnitudes
are often $\sim$0.2 brighter than V magnitudes. Exceptions to this
general rule include FO~Aqr, which experienced a deep low state
in 2016 \citep{Littlefield2020}, and this appears to be biasing
the average \gaia\ value lower (G=13.9) compared with V$\sim$13.7
that is typical of its normal state. Similarly, the \gaia\ average
for V1025~Cen may be biased by as much as $\sim$1.5 mag lower due
to a low state that is seen in the ASAS-SN data (\citealt{Covington2022};
note that there are no AAVSO data for this IP during the relevant period).
GK~Per, on the other hand, had an outburst in 2015, which appears
to have biased the average \gaia\ measurement higher (G=12.6) compared
to typical quiescent V magnitude of V$\sim$13.1. However, this is
because the outburst of GK~Per is exceptionally long (see, e.g.,
\citealt{Hameury2017}); most IP outbursts are too infrequent
and too brief to bias the \gaia\ averages significantly.  We therefore
take the \gaia\ G magnitudes as representative of their normal state,
for the vast majority of these IPs.

The absolute G magnitudes of these 67 IPs range from 2.9 to 11.6, with
a majority (37 systems) having \gabs\ in the 4.0--6.0 range. A simple
histogram of IPs by their \gabs\ will have a single broad peak in this
range, with five systems on the bright end and a long tail of 25 systems
on the fainter end. While this is not particularly informative, a clear
picture emerges when IPs are placed in the orbital period vs. absolute
G magnitude plane (Figure\,\ref{fig:llipob}). In this figure, 61 out of
67 IPs for which the orbital period is securely known are plotted.

The vast majority of these IPs fit within the standard
framework of CV evolution driven by angular momentum loss \citep{Knigge2011}.
In these systems, the secondary is on or near the main
sequence, whose luminosity can be estimated, to first order, from the orbital
period alone.  The secular-average accretion rate can also be estimated from
the orbital period, but with considerable scatter. In contrast, 5 of the 67
IPs have orbital periods in excess of 10 hrs (GK~Per, Swift~J2006.4+3645,
V2731~Oph, IGR~J17014-4306, and RX~J2015.6+3711 in order of decreasing \porb).
The G magnitudes of these long-period IPs have a strong contribution from
the evolved secondary, and the secular accretion rate is even more uncertain.
The number of known CVs, magnetic or otherwise, with \porb\ $>$10 hrs
is relatively low and they exhibit a wide range of absolute G magnitudes
and outburst behaviors. We therefore concentrate on IPs with \porb\ $<$10 hrs
in the rest of this paper.

\subsection{Disc Brightness Index}

\begin{figure*}
\centering
\includegraphics[width=5.0 in]{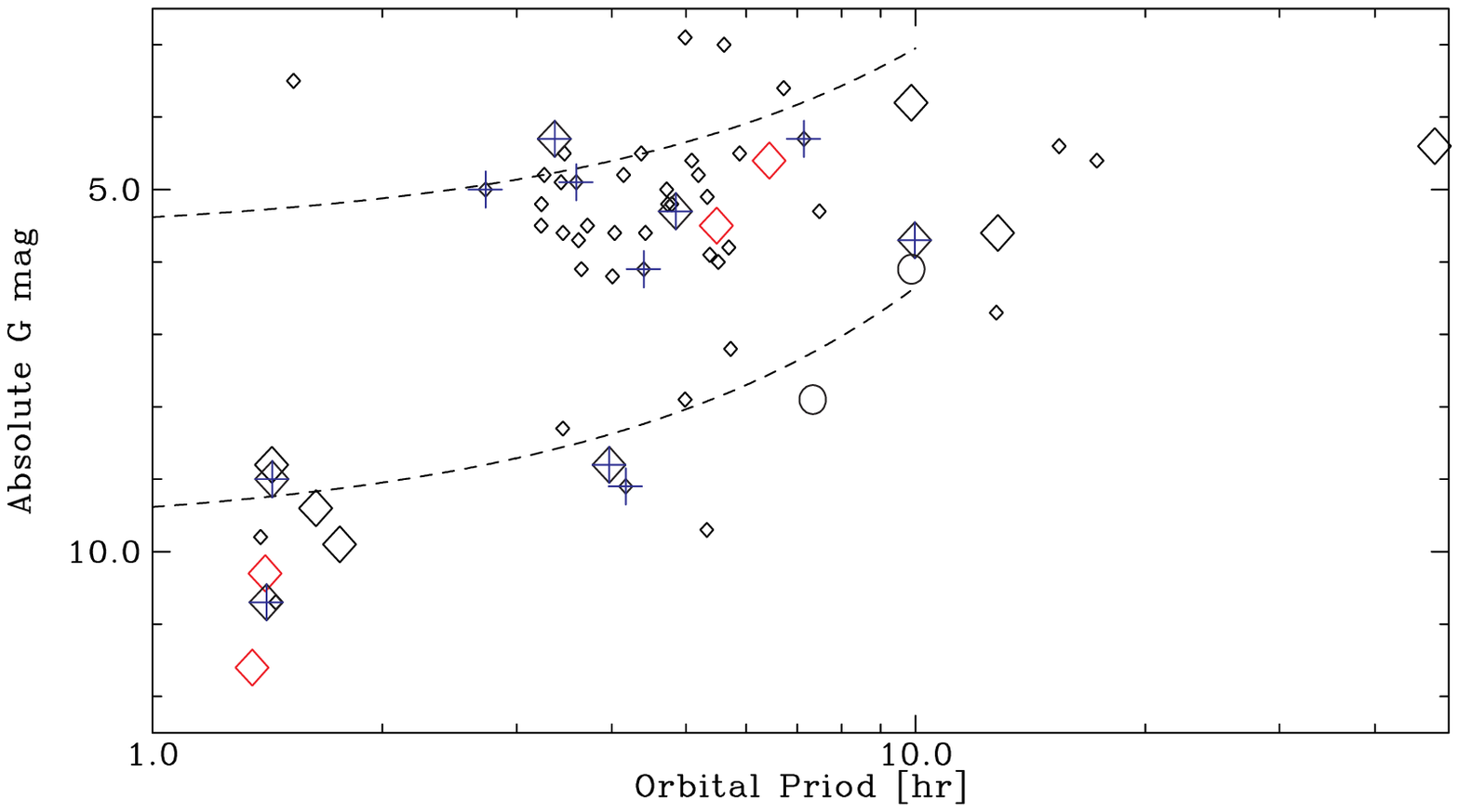}
\caption{The absolute G magnitudes of 61 IPs are shown against their orbital
periods. The two candidate propellers, AE~Aqr and LAMOST J024048.51+195226.9,
are plotted as circles. The size of the diamonds indicates whether outbursts
are known (large) or not (small). Four IPs in which positive superhumps have been seen are shown in red.
IPs with known low states are
indicated by large blue plus signs overplotted on the diamonds.  
The dashed curves show approximate locations of dwarf novae during quiescence and in outburst.}
\label{fig:llipob}
\end{figure*}

In Figure\,\ref{fig:llipob}, we also plot curves
representing fits to the absolute magnitudes of dwarf novae in outburst
and in quiescence. Dwarf novae are a common and well-studied type of
non-magnetic CVs that are seen in the faint, quiescent, state the majority
of the time, punctuated by occasional outbursts.  Specifically in
Figure\,\ref{fig:llipob}, we use the absolute V magnitude-\porb\
relationship for quiescent dwarf novae (\qmag =9.72$-$0.337\porb)
and those in outburst (\omag =5.64$-$0.259\porb) according to
\cite{Warner1987}. Note that these relationships are only calibrated
for \porb\ $<$10 hrs.

The outburst absolute magnitudes of dwarf novae, as a function of \porb,
can be explained as due to high state discs having near-constant surface
brightness, and the \porb\ dependence of
the radius of the primary Roche lobe, and hence the accretion disc
\citep{Warner1987}. The case of dwarf novae in quiescence is far more
complicated, with contributions from the emission lines, the secondary
in the long orbital period systems, and the white dwarf in particularly faint systems. Nevertheless, it is not surprising that there is an orbital period dependence for quiescent dwarf novae. Similarly, \gabs\ of IPs, too, are expected to depend, in part, on \porb.

There are several caveats and limitations for the comparison between IPs and Warner's relationships for dwarf novae. First, \cite{Warner1987} used the distances to the dwarf novae that were estimated using the techniques available at the time. Compared to what could be done with \gaia\ data, there could be additional scatters or offsets in the data used to construct these relationships.
Second, Warner's formula refer to the V band magnitudes, not the broad (effectively white light) \gaia\ G band. Dwarf novae at maximum are dominated by the hot discs and tend to have B$-$V$\sim$0.0, V$-$R$\sim$0.0, and R$-$I$\sim$0.0 (see, e.g., \citealt{Spogli1998}) so \omag\ should be a reasonable approximation for the
equivalent absolute G magnitude M$_{\rm G,o}$. However, the conversion between V and G magnitudes for quiescent dwarf novae is much less certain, and this is a potential source of errors of our study.
Note also that we have not corrected \gabs\ of IPs for interstellar reddening, since there
is no database of uniform and reliable reddening data for the IPs.  Finally, inclination angles are poorly known for most IPs, so we have not corrected \gabs\ of IPs for inclination angle effects. This was done for dwarf novae by \cite{Warner1987}: discs seen face-on can be 1 magnitude brighter than average, while edge-on discs of deeply eclipsing systems are fainter by 1.7 magnitudes or more \citep{Patterson2011}.
Nevertheless, Warner's lines probably should give a reasonable indication
of whether an IP is similar in the optical to dwarf novae during quiescence
or in outburst.

\begin{figure}
\centering
\includegraphics[width=3 in]{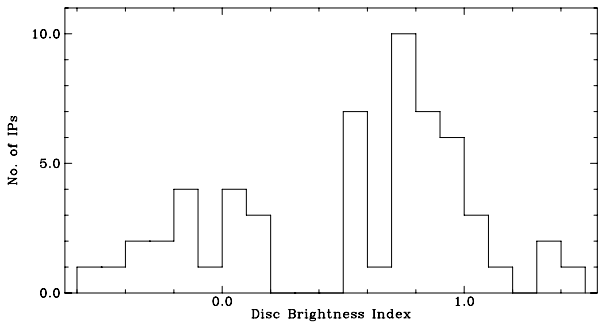}
\caption{The distribution of IPs with \porb\ less than 10 hrs as a
function of their disc brightness index.}
\label{fig:dbihisto}
\end{figure}

Figure\,\ref{fig:llipob} gives the impression of a dichotomy among IPs
with \porb\ below 10 hrs, one group near the quiescent dwarf nova line
and the other near the outburst line. In an attempt to quantify this,
we define a new parameter, which we call the disc brightness index (DBI), as
[\gabs $-$ \qmag]/[\omag $-$ \qmag] where \qmag\ and \omag\ are
functions of \porb, as prescribed by \cite{Warner1987}.
By construction, DBI of 0.0 indicates an object similar to a quiescent
dwarf nova with the same orbital period, and 1.0 an object similar to
dwarf nova with the same orbital period in outburst.  The DBI values
of the IPs are tabulated in Table\,\ref{tab:ipabsmag}, and a histogram
of the number of IPs as a function of DBI in 0.1 increment is shown
in Figure\,\ref{fig:dbihisto}.   Note that the two
relationships of \cite{Warner1987} differ by $>$3 magnitudes at all
orbital periods for which they are defined.  Therefore, an accuracy
in DBI of 0.1 requires that \gabs\ of IPs to be known to $\sim$0.3 mag,
which appears to be achieved for the majority of IPs in
Table\,\ref{tab:ipabsmag}.  One exception is V1025~Cen: if its low state
resulted an underestimate of its \gabs\ by 1.5 mag, then its DBI is
underestimated by 0.379, implying that it may have a DBI of +0.012 in
its normal state. Since the zero point of the formula to correct for
the inclination dependence of disc brightness (equation 4) used by
\cite{Warner1987} results in an average correction of $-$0.367 mag,
the DBI of disc-dominated IPs are likely underestimated by $\sim$0.1.

This figure appears to confirm the dichotomy of IPs into two subclasses.
Of the 56 IPs for which we have calculated the DBI (61 objects included
in Figure\,\ref{fig:llipob} minus 5 with \porb\ $>$10 hrs), 18 have DBI
less than 0.2, 38 have DBI over 0.5, and there are none in the 0.2--0.5
range.  We will use the term optically-determined low luminosity IPs
(LLIPs) for the former, and optically-determined high luminosity IPs
(HLIPs) for the latter. Note that, in the context of this work, these terms are not applied to IPs with orbital periods longer than 10 hrs.

Figure\,\ref{fig:dbihisto} should not be taken as a true representation
of the underlying population. In particular, selection effects against
discovering and/or recognizing LLIPs as such are likely. However, it
would take a contrived set of circumstances for selection effects to
create the appearance of a gap at DBI range of 0.2--0.5 if there was
no such gap in the underlying population. Similarly, the various caveats
mentioned above for the use of Warner's lines may shift the peaks or
broaden the distributions in DBI, but we cannot come up with a plausible
scenario in which one or more of these limitation would create an apparent
DBI gap at 0.2--0.5. Therefore, the conclusion that there are two separate
populations seems reasonably secure.

\subsection{Comparison of Optical and X-ray Luminosities}

This is not the first time that high- and low-luminosity subclasses of IPs were considered.
Previously, \cite{Pretorius2014} found that, while most IPs had hard X-ray
(BAT-band) luminosity higher than $\sim 10^{33}$ \eps, there was an
indication of a separate population of LLIPs with BAT-band luminosity
of order $10^{31}$ \eps.

\begin{figure}
\centering
\includegraphics[width=3 in]{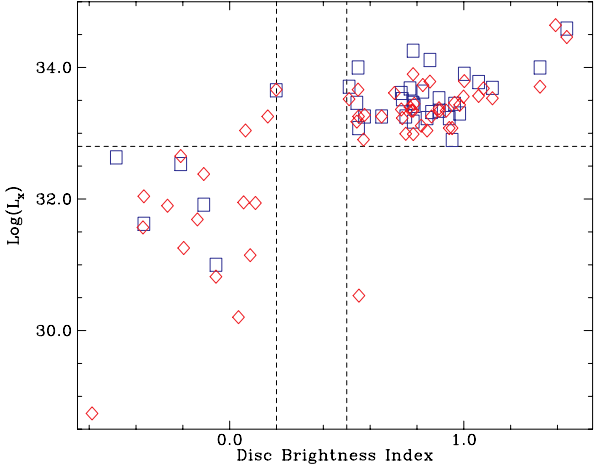}
\caption{Hard (blue square) and soft (red diamond) X-ray luminosities
of IPs, as a function of their disc brightness index (DBI). Vertical dashed
lines are drawn at DBI=0.2 and 0.5, and a horizontal dashed line is drawn
at log(L$_X$)=32.8.}
\label{fig:dbilx}
\end{figure}

To investigate anew the X-ray based distinction between HLIPs and
LLIPs, we have also collected hard X-ray fluxes from the \swift\ BAT 157
month survey\footnote{https://swift.gsfc.nasa.gov/results/bs157mon/}
(see details of its predecessor, the BAT 105 month survey, in
\citealt{BAT105}), and the ``soft'' ($<$10 keV) X-ray fluxes primarily
from the fourth \xmm\ serendipitous source catalogue (4XMM-DR11;
\citealt{Webb2020}) and the \swift\ X-Ray Telescope Point-source Catalog
(2SXPS; \citealt{Evans2020}). Neither catalog has an entry for CC Scl,
but it is listed in the \xmm\ Slew Survey Catalog \citep{XMMslew}.
Pointed \xmm\ and \swift\ observations of GK~Per are taken during
its outbursts, so we used the quiescent \asca\ observation \citep{Ezuka1999}
instead for this object. For V4743~Sgr (=nova Sagittarii 2002 No. 3), we
selected the 4XMM-DR11 entry for observation dated 2006 March 28, as this
observation is most likely to be dominated by the accretion luminosity,
rather than the residual effects of nova eruption as may well be the case
for earlier observations.

All in all, we have collected 42 hard X-ray fluxes and 65 soft X-ray fluxes
for the 67 IPs, and converted them to luminosities, with only one object,
the candidate propeller LAMOST~J024048.51+195226.9, without either.
Of the 56 IPs for which DBI can be defined, hard X-ray luminosities
are known for 35 objects and soft X-ray luminosities for 54 objects.
These are plotted against DBI in Figure\,\ref{fig:dbilx}.  We also
plot a horizontal dashed line at log(L$_X$)=32.8 to guide the eye.

It is clear that the optically-defined HLIPs are almost always X-ray
luminous at this level, with the exception of DQ~Her (DBI=0.553), which
is not detected with BAT and has an \xmm\ luminosity of 3.4$\times$10$^{30}$
\eps. However, DQ~Her is a deeply eclipsing system and it is thought
that the the viewing geometry is so exactly edge-on that we have no
direct line of sight to the post-shock region
above the white dwarf surface \citep{Mukai2003}.
The intrinsic X-ray luminosity of DQ~Her is therefore likely much higher
than the observed luminosity.  Note also that, when corrected for the
high inclination, the DBI of DQ~Her is also significantly higher than
the value adopted here (0.553).
%Two other HLIPs nearest the the log(L$_X$)=32.8
%line are V1323~Her (DBI=0.572, no BAT detection, \xmm\ luminosity of
%8.0$\times$10$^{32}$ \eps) and AO~Psc (DBI=0.950, BAT luminosity of
%7.9$\times$10$^{32}$ \eps\ and \xmm\ luminosity of 1.2$\times$10$^{33}$
%\eps).

In contrast, the optically-defined LLIPs are widely distributed
in their X-ray luminosities but predominantly below the log(L$_X$)=32.8
line. The exceptions are V1062~Tau (DBI=0.199, BAT luminosity of
4.5$\times$10$^{33}$ \eps\ and \swift\ luminosity of 4.6$\times$10$^{33}$
\eps), V3037~Oph (DBI=0.163, no BAT detection, and \xmm\ luminosity of
1.8$\times$10$^{33}$ \eps), and IGR~J19267+1325 (DBI=0.068, no BAT detection,
and \swift\ luminosity of 1.1$\times$10$^{33}$ \eps). Of
these, V1062~Tau has an orbital period of 9.98 hr, at the edge of the
regime where DBI can be defined.  Perhaps our analysis based on DBI is
not valid for this system. The latter two may suffer substantial optical
reddening (A$_V \sim$2 and A$_V$=1.4$\pm$0.7 respectively;
\citealt{Thorstensen2013,Butler2009}).  If verified, this could push
their DBI numbers high enough to place them in the 0.2--0.5 gap or even
among the (optically-determined) HLIPs.  This does point out the need
for a systematic investigation of interstellar reddening to see if the
gap between HLIPs and LLIPs (Figure\,\ref{fig:dbihisto}) is complete
or if there are a few in-between systems. 

%The remaining LLIPs range from V455~And
%(DBI=$-$0.588, no BAT detection, \xmm\ luminosity of 5.5$\times$10$^{28}$
%\eps) to 1RXS J211336.1+542226 (DBI=$-$0.209, BAT luminosity of
%3.4$\times$10$^{32}$ \eps\ and \xmm\ luminosity of 4.5$\times$10$^{32}$ \eps)
%and PBC~J1841.1+0138 (DBI=$-$0.484, BAT luminosity of
%4.3$\times$10$^{32}$ \eps\ and no soft X-ray catalog entry).

We can conclude that the X-ray luminosity and the optical
brightnesses, as indicated by DBI, of IPs are
strongly correlated, with possible outliers. Both are therefore likely
to be controlled largely by a single parameter, presumably the accretion
rate. In contrast, \gabs\ of IPs does not provide
a useful discriminant, as it is also strongly influenced by the
size of the partial accretion disc.

\section{Discussion}

\subsection{IP Subclasses and Orbital Periods}
\label{LPIP}

%Long period (\porb\ $>$ 10 hr) IPs have large discs. Given the radial
%dependence of the critical accretion rate for the disc to be in the
%high state (see, e.g., equation 1 of  \citealt{Schreiber2007}), parts
%or all of discs of long period IPs may remain in low state even when
%the mass transfer rate is high.  

%Now we return to the subject of the division between HLIPs and LLIPs,
%among systems with \porb\ $<$ 10 hr.
It is well known that CVs in general have lower mass transfer rates below
the 2--3 hr ``period gap'' than above \citep{Knigge2011}, and this
indeed turns out to be a important factor in determining which IPs have low
X-ray and optical luminosities. Specifically, 9 out of 10 IPs below
the period gap are seen to be LLIPs. These systems are relatively nearby, and so a large correction due to interstellar reddening
is not expected for these systems.

The 10th system, IGR~J18173$-$2509, was suggested to be an LLIP based
solely on its short orbital period \citep{Mukai2017}. However, it
is luminous both in X-rays ($> 10^{34}$ \eps) and in the optical
(\gabs=3.5, DBI=1.440). These numbers were calculated
using a distance of 4.69 kpc.  However, the the \gaia\ parallax of
0.2212$\pm$0.0859 mas corresponds to a possible range of 3.62--6.14 kpc
according to \cite{GaiaEDR3}. Using the lower bound instead,
IGR~J18173$-$2509 has \gabs=4.1, or DB=1.275, so the distance uncertainty
does not impact our assessment that this system is an HLIP.  Therefore,
it must have an accretion rate far in excess of the secular average,
perhaps similar to (but not as extreme as) T~Pyx
(\porb =1.83 hr, M$_V$=0.9 according to \citealt{Patterson2017},
although uncertainties remain regarding the distance to this object).

In contrast, in the orbital period range of 2.7 to 10 hrs, there are 9
LLIPs and 37 HLIPs.  Of the 9 LLIPs in this range, two systems,
AE~Aqr and LAMOST~J024048.51+195226.9, are candidate propeller
systems (see, e.g.,
\citealt{Thorstensen2020,Pretorius2021} and references therein).
As already mentioned, some IPs in this period range (e.g., V3037~Oph
and IGR~J19267+1325) may have been misclassified as LLIPs due to large
interstellar extinction.  However, there are IPs in this orbital period
range that appear to be genuine LLIPs (e.g., DO~Dra and V1460~Her),
despite the expectations from considerations of secular accretion rate.

LLIPs are sufficiently optically faint to preclude the presence of
high state discs.  The optical brightness of HLIPs suggests the
presence of high state discs, and certainly allows it, even though
there are obvious factors that could limit the validity of any direct
comparisons of properties of discs in non-magnetic CVs and IPs.
The latter should have a central hole, so the total disc area is
somewhat smaller than that in a non-magnetic system. On the other hand,
irradiation of the partial disc and the accretion curtains should add
optical light beyond the disc luminosity provided by the in-situ viscous
heating, as evidenced by the presence of optical spin modulation.
Interestingly, LLIPs with small \porb\ / \pspin\ ratios
(such as EX~Hya, \porb\ / \pspin\ = 1.4659) and large ratios
(CTCV~J2056$-$3014, 322) appear to have similar \gabs, despite the
expected differences in the size of the central hole in their discs.
Another interesting case is that of V2400~Oph. Even though it was originally
established to be a discless IP \citep{Buckley1995, Buckley1997},
it is nevertheless an HLIP according to this analysis. Note also that recent
X-ray observations have led \cite{Joshi2019} to reconsider the
possible presence of a disc, or disc-like structures, in V2400~Oph,
at least during certain epochs.

HLIPs, LLIPs, and long-period IPs also have distinct
characteristics in their optical spectra. The continuum shows strong
contributions of the mass donor in long-period IPs, and any disc
contributions can only be determined after a careful analysis. Only
the most prominent emission lines are obvious above the bright continuum.
Of the 4 IPs shown in Figure\,2 of \cite{Gaensicke2005}, V2731~Oph
(=1RXS~J173021.5$-$055933, \porb=15.4 hr) is a good example of this.
HLIPs generally show blue non-stellar continuum from the disc,
similar to those of dwarf novae in outburst, on which prominent emission
lines of Hydrogen Balmer series as well as HeII $\lambda$4686 are
superimposed.  The strengths of these lines distinguishes HLIPs from
non-magnetic CVs with bright discs. In Figure\,2 of \cite{Gaensicke2005},
V647~Aur (=RXS~J063631.9+353537) and V418~Gem (=1RXS~J070407.9+262501)
are typical of this.  Interestingly, the final panel of this figure
(for V1323~Her=1RXS~J180340.0+401214) shows a faint, flat continuum
with strong emission lines (including Balmer jump in emission) that is
a hallmark of low state discs.  The only clue that this system may be
magnetic is the relative prominence of the HeII $\lambda$4686 line.
In this respect, this spectrum of V1323~Her is similar to optical
spectra of LLIPs.  The fact that this system is among the lowest
luminosity HLIP (DBI=0.572, no BAT detection, \xmm\ luminosity
of 8.0$\times$10$^{32}$ \eps) might be related to this.

%Porb uncertain - 6
%WX Pyx (5.3)
%IGR J08390 (8) - LLIP?
%IGR J04571 (6.2 or 4.8) Either
%SWfit J183920.1-045350 (5.6) - LLIP?
%AX J1740.1-2847 (2.1) - LLIP
%AX J1832.3-0840 (1.32 ?) - LLIP

In the following subsections, we review selected
results on IPs in recent publications, obtained using time-domain
observations. Our discussion is informed by the distinction between
HLIPs and LLIPs.

\subsection{Superhumps in IPs}
\label{IPSH}

%Given the questions regarding both the magnetically gated accretion
%model V1223~Sgr \citep{Hameury2022} and the micronova model
%\citep{Scaringi2022Nature,Scaringi2022MNRAS},it seems prudent to
%explore possible alternatives. In this subsubsection, however, we take
%a detour to discuss superhumps in IPs to set up the
%model to be proposed in the next subsubsection.

Superhumps are defined as photometric variations at periods
slightly different from the orbital period. Many are at periods
slightly longer (positive superhumps), while some have periods
that are slightly shorter (negative superhumps). The former is
generally believed to be due to apsidal motion
of elliptical discs \citep{Whitehurst1988,Osaki2005}, and the
latter due to nodal precession of tilted discs \citep{Wood2000,Thomas2015}.
%These phenomena reveal the limitation of one-dimensional (vertically
%integrated, azimuthally averaged) analyses of accretion discs,
%however useful they may be. In particular, the HLIP, \tvcol,
%is well known for its negative superhump (5.2 hr period, vs.
%\porb=5.5 hr, with a 4-day disc precession period; \citealt{Barrett1988}),
%so a full understanding of its partial disc requires 3-dimensional
%studies.

Positive superhumps were first observed during superoutbursts
of SU~UMa-type dwarf novae that are largely found below the period
gap. The definition of this subclass is that they have superoutbursts
and normal outbursts. Superoutbursts, in turn, are longer and brighter
than normal outbursts, and are defined by the presence of superhumps
\citep{Osaki2005}.  When the outer radius of the disc grows beyond
the 3:1 resonance radius, the disc becomes elliptical and starts to
precess; when the outermost region of the elliptical disc is nearest
the secondary, there is enhanced tidal stress, which results in increased
disc brightness. Tidal truncation, however, limits the outer radius
of the disc \citep{Paczynski1977}; since this is a function of the
mass ratio $q$ (secondary mass divided by the white dwarf mass),
superhumps are expected only for systems with certain values of $q$.
Early theoretical calculations suggested $q <$0.25 as the precondition
for superhumps \citep{Whitehurst1988}, requiring low mass secondaries
and hence short orbital periods. Superhumps are also
seen in novalike systems (non-magnetic CVs that are persistently
bright, comparable to dwarf novae in outburst) and these are often
referred to as permanent superhumpers.

Observationally, \cite{Patterson2005} found that confirmed
superhumps in non-magnetic CVs are limited to systems with \porb\ $<$3.5 hrs
and $q$ smaller than 0.35$\pm$0.02.  However, \cite{Retter2003} reported
the detection of a 6.3 hr positive superhump (in addition to the negative
superhump that is usually present) in the HLIP, \tvcol. While this has
not been confirmed in \tess\ data \citep{Scaringi2022Nature,Bruch2022},
the former reports the discovery of a $\sim$7.8 hr superhump period in
\eiuma\ (\porb =6.435 hr) instead. Thus, \eiuma\ now holds the record as
CV with the longest orbital period for which positive superhumps have
been detected, indirectly enhancing the credibility of the purported
superhump detection for \tvcol\ \citep{Retter2003}.  Note that the search
for transient positive superhumps in \tvcol\ is a challenging task even
using \tess\ data, given the persistent presence of the orbital period
and the negative superhumps, particularly when the system brightness
is changing on a comparable timescale.

In Figure\,\ref{fig:llipob}, we used red symbols to indicate four IPs
for which positive superhumps have been reported. Two of them, V455~And
\citep{Matsui2009} and CC~Scl \citep{Woudt2012}, are systems below the
period gap that have shown superoutbursts. Presumably,
the discs in these LLIPs can reach the 3:1 resonance radius without
exceeding the tidal truncation radius, so no special explanations
are necessary. The other two are \tvcol\ and \eiuma.

This is a surprising result, because superhumps
are not expected according to the standard model in such long
period systems.  It is perhaps worth considering if this is
related to the truncated nature of IP discs. It is also worth
noting that superhumps in these HLIPs appear not to be persistent
features of these systems.

\subsection{Implications for the Low States of IPs}

It is now apparent that low states of IPs are not as
rare as previously thought (see \citealt{Covington2022}
and references therein).  Those with known low states
are noted in Table\,\ref{tab:ipabsmag} and shown with
an added plus symbol in Figure\,\ref{fig:llipob}.
It is also clear that low states are seen in HLIPs and LLIPs.
However, not all low states are equal.  Observers generally
define low states in terms of relative photometry --- for example,
\cite{Covington2022} used ``a sustained drop in flux of $\gtrsim$0.5 mag''
as their definition. What a given drop in optical brightness implies
depends on the absolute magnitude in the normal state and
the contribution of the secondary to it.

%In the long-period IPs, the donor is evolved and luminous
%enough to contribute a significant fraction of optical
%brightness in their normal state, depending also on the
%bandpass of choice. For this reason alone, fluctuations in
%accretion rate will be harder to detect through optical
%observations. This is consistent with the fact that no
%low states have been reported among the 5 long-period IPs
%so far, although this may well be in part due to the small
%number of systems involved.

The accretion luminosity of an LLIP is modest even during
the normal state.  Therefore, when an LLIP goes into a
low state, it can easily enter a state in which optical light
is dominated by the component stars, with little or no contribution
from accretion.  For example, in DO~Dra (M$_G$=8.8 in \gaia\ EDR3),
the M dwarf mass donor is detectable in the red part of the spectrum
\citep{Mukai1990}, even in its normal state. The deepest part
($\sim$2 mag in the V band) of its low state is interpreted by
\cite{Covington2022} as due to cessation of accretion onto the
white dwarf. This is corroborated by the non-detection in X-rays
during a 55 ks \nustar\ observation \citep{Shaw2020}, and by a
low-state \tess\ light curve in which only the orbital
modulation (via the ellipsoidal modulation) was detectable
\citep{Hill2022}.  Similarly, accretion may have completely
stopped in V1025~Cen at its lowest flux \citep{Covington2022}.

\begin{figure}
\centering
\includegraphics[width=3 in]{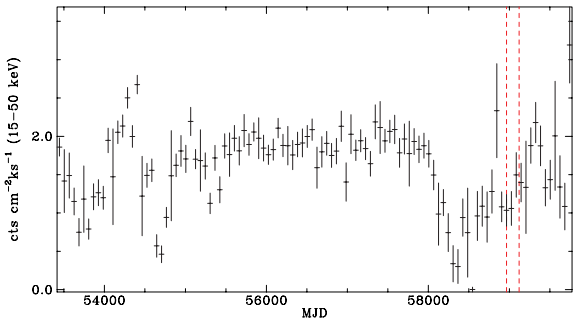}
\caption{\swift\ /BAT light curve of V1223~Sgr, binned into 60 day bins
and plotted in units of cts\,cm$^{-2}$ks$^{-1}$. There is one bin, centred
on MJD 58545 (2019 March 3) that is statistically consistent with 0
($1.8\times 10^{-6} \pm 3.1 \times 10^{-5}$); otherwise V1223~Sgr
is always detected. Vertical dashed line indicates time intervals during
which short bursts were seen.}
\label{fig:v1223bat}
\end{figure}

On the other hand, the observed low states of HLIPs are generally
a move towards the LLIP level of optical brightness.
For example, the low state of V1223~Sgr \citep{Covington2022}
was a drop of roughly 2 magnitudes to an absolute magnitude
of $\sim$6.3 (or DBI$\sim$0.6). The \swift\ BAT
%assuming that the \gaia\ G magnitude is representative of the normal state.
data on V1223~Sgr plotted using 60 day-bins (Figure\,\ref{fig:v1223bat};
obtained from the \swift\ BAT transient monitor
site\footnote{https://swift.gsfc.nasa.gov/results/transients/};
\citealt{Krimm2013}) suggests accretion was on-going through much
of the low state, with the hard X-ray fluxes dropping by a factor
of 2--5 from the long-term average.  We do note that there is one bin which
is significantly lower. Similarly, the study of archival
photographic plates by \cite{Garnavich1988} indicates a couple
of epochs of non-detection in the optical. The upper limit is 3.8 mag below
the typical high state magnitudes, suggestive of absolute
magnitude $>$8.1. At these epochs, V1223~Sgr may have been
comparable to LLIPs, or it may have been fainter.

In all, there does not appear to be a secure detection of
an HLIP in which accretion has temporarily ceased. Moreover,
there do not appear to be high time resolution time-series
photometry, optical spectroscopy, or pointed X-ray observations
of an HLIP in a deep enough low state to temporarily qualify as
an LLIP (or fainter).  What we do have are such observations during shallow
low states in which the objects is still brighter than typical
LLIPs (e.g., FO~Aqr; see \citealt{Kennedy2017,Littlefield2020}).
It presumably requires active accretion to keep the disc bright,
either via in-situ viscous heating or via irradiation from
accretion onto the white dwarf.

% Ava's LLIP: DW Cnc (delta~2 mag), DO Dra, V1025 Cen (delta~2 mag)
% Ava's HLIP: V515 And (delta~1 mag), V1223 Sgr (delta~2 mag),
%             RX J2133.7+5107 (delta~1.5 mag)

To recap, low states have been observed in both HLIPs and LLIPs.
% but not (so far) in long orbital period IPs.
These low states presumably result from downward fluctuations in
the mass transfer rate from the secondary. However, not all low
states are equal, which should be kept in mind in interpreting
the low state data, as well as in considering the physical origin
of the mass transfer rate variability. In particular,
we do not yet have contemporaneous observations of HLIPs in a deep
low state in which accretion has ceased.  Continued monitoring of
HLIPs is required to see if they ever enter such deep low state
and, if one does, to observe it in detail.

%BAT light curve of V1223 Sgr.
%2459000 = 2020-05-30
%8970-9150, so recovery was from 2020 May to October
%(including Chandra HETG obs)

\subsection{Implications for the Outbursts of IPs}

In recent years, an increased attention has been paid to
outbursting behaviour of IPs. We list the IPs with known outbursts
in Table\,\ref{tab:ipabsmag} based on the list provided in
\cite{Hameury2017}, supplemented with additional systems noted
by \cite{Hameury2022}, as well as noting papers on individual
objects (V1025~Cen; \citealt{Littlefield2022} and IGR~J17014$-$4306;
\citealt{Shara2017}). IPs with known outbursts are plotted using
larger diamonds in Figure\,\ref{fig:llipob}.

Dwarf novae are considered to be non-magnetic systems (i.e., any
magnetic field the white dwarf might possess is not strong enough to
influence the accretion flow) in which the mass transfer rate from
the secondary to the disc is low. Dwarf nova outbursts are
interpreted as due to thermal instability of the accretion discs
\citep{Lasota2001}. In quiescence, the disc is in a cool, faint state
and the disc mass gradually grows; when the surface density
becomes sufficiently high, it transitions to a hot, bright
state (outburst).  The same basic mechanism can in principle operate
in LLIPs with magnetically truncated discs, with similar or somewhat
modified outburst properties (such as recurrence period and duration).

Indeed, of the 18 LLIPs as identified by DBI, 9 are known to have
outbursts.
%, including the borderline long-period IP, V1062~Tau.
Their outburst durations (see, e.g., Table~1 of \citealt{Hameury2017})
are generally longer than 1 day, and these outbursts can
plausibly interpreted as to due to thermal instability of the disc,
under the limited influence of the magnetic field
\citep{Hameury2022}. Interestingly, there are hints that
the size of the central hole (using the \porb\ / \pspin\ ratio
as proxy) might influence the outburst duration and amplitude.
For example, outbursts of EX~Hya (\porb\ / \pspin\ = 1.4659)
is a relatively short ($\sim$2 day) and their amplitude modest ($\sim$3.5 mag;
\citealt{Hellier1989}). In contrast, the 2007 superoutburst
of V455~And (\porb\ / \pspin\ = 71.9482) lasted well over 20 days
and had an amplitude greater than 7 mag \citep{Matsui2009}.
The only exception among LLIPs is the short, repeated bursts seen
in V1025~Cen \citep{Littlefield2022}, which the authors interpreted
in the framework of accretion gating model.

On the other extreme, GK~Per exhibits $\sim$2 month-long
outbursts \citep{Evans2009}. This is presumably a consequence of the
large physical size of the disc in this system, which has an
exceptionally long, $\sim$2-day orbital period. The outbursts seen
in archival photographic plates in IGR~J17014$-$4306 \citep{Shara2017}
may turn out to be similar.
%In addition to the long duration,
%the large size of the discs may allow the outer regions to remain
%cool enough for thermal instability to operate, even while the white
%dwarf is actively accreting \citep{Schreiber2007}.

HLIPs, on the other hand, are likely to have bright discs that are
stable against dwarf nova outburst. It is true that the absolute magnitudes
of HLIPs often appear fainter than those of dwarf novae at maximum
(Figure\,\ref{fig:llipob}), and hence their DBIs are on average lower than
1.0 (Figure\,\ref{fig:dbihisto}). However, the critical value for stability
is thought to be smaller than the maximum brightness of dwarf nova
outbursts, which is confirmed by the standstills of Z~Cam type dwarf
novae roughly 0.7 mag below outburst peak \citep{Lasota2001}.

The outbursts that require a radical new interpretation therefore are the
short events seen in HLIPs \tvcol, \eiuma, V1223~Sgr, NY~Lup,
and FO~Aqr \citep{Scaringi2022Nature,Hameury2022}. One such
radical interpretation is the micronova model of
\cite{Scaringi2022Nature,Scaringi2022MNRAS}, which
proposes that local thermo-nuclear runaway (TNR) can happen on
magnetic CVs, as opposed to the well-established phenomenon of
classical nova eruptions (see \citealt{Chomiuk2021} for a recent
review), which is presumed to be a phenomenon over the entire
surface of the accreting white dwarf. There are
missing elements in the current literature of micronovae that
need to be filled in, if this model is to be widely accepted.
One is the process in which the energy of the local TNR is converted
to visible light. Another is a quantitative assessment of whether
the white dwarf magnetic field is strong enough to keep the
plasma confined at near TNR conditions, analogously to what was
done by \cite{Hmeury1983} in the neutron star case.

Another interpretation is the magnetically gated accretion model,
which has been applied both to the LLIP V1025~Cen
\citep{Littlefield2022} and to HLIP V1223~Sgr \citep{Hameury2022}.
At the time when repeated short outbursts are seen, V1223~Sgr was
in a shallow low state \citep{Hameury2022} that likely implies that
it had a bright partial disc. There is no quantitative exploration
in the literature yet of how magnetically gated accretion can lead
to enhanced optical emission sufficiently luminous to compete with
the partial disc in HLIPs.

%Further exploration is necessary to determine if
%a single model can apply to both these cases.

Observations suggests that the short outbursts of
HLIPs and superhumps may be linked. Two HLIPs with positive
superhumps are both systems known to have short outbursts.
Furthermore, \cite{Retter2003} noted that the superhumps were
observed right after the 2001 January 7 mini-outburst (their
terminology) of \tvcol, but not immediately before. In the case
of \eiuma, \cite{Scaringi2022Nature} noted that the superhumps
were observed before the bursts, but not after. We do not know
yet if superhumps and  short outbursts are causally connected,
but we believe that this is a possibility worth exploring.

Note that, in SU~UMa-type dwarf novae, superoutbursts
are characterized by both the presence of superhumps and by the
higher luminosities in comparison to normal outburst.  In one model,
the thermal-tidal instability (TTI; \citealt{Osaki1989}),
the tidal instability of an elliptical disc is the cause not only
of the superhumps but also of the enhanced accretion luminosity,
although in another, tidal interactions are merely a consequence
of superoutbursts \citep{Cannizzo2012}, Regardless of the which of these
interpretations is correct, enhanced luminosity and the presence of
superhumps are physically linked in SU~UMa type dwarf novae. Thus, the
contemporaneous presence of superhumps and short outbursts in
HLIPs TV~Col and EI~UMa is highly suggestive.

\section{Conclusions}

We have compiled \gabs\ of confirmed IPs, and plotted against their
orbital period.

\begin{enumerate}

\item IPs with orbital periods below $\sim$10 hrs are separated into
an optically bright subclass (HLIPs), and an optically faint subclass (LLIPs).
These subclasses are strongly correlated with the X-ray luminosity-based
subclasses originally proposed by \cite{Pretorius2014}.

\item Low states in LLIPs often are significant events, some being
consistent with a complete cessation of accretion. In contrast,
low states in HLIPs are often modest fluctuation within the HLIP
regime.

\item Of all the outbursts in IPs, short outbursts
in HLIPs are the most challenging to interpret. So are the transient
superhumps observed in HLIPs TV~Col and EI~UMa. The contemporaneous
detection of short outbursts and superhumps in these systems
suggests the possibility that these two share a common physical origin.
We propose that searches be conducted for transient positive superhumps
in all HLIPs that have shown short outbursts.
\end{enumerate}

\section*{Acknowledgments}

We fondly remember our friend, mentor, and colleague, Brian Warner, and thank
him for intellectual and personal guidance.  While his passing leaves a huge
hole in our field, his body of work will continue to inform and inspire us
for many years to come.
We also thank Dr. Ken Shen for stimulating conversations and for
his careful reading of a draft version of this paper.
We acknowledge with thanks the variable star observations
from the AAVSO International Database contributed by observers worldwide
and used in this research.

\section*{Data Availability}

The \gaia\ data presented in this article are available at
ESA \gaia\ archive (https://gea.esac.esa.int/archive/).
The X-ray flux data presented in this article are available
at the High Energy Astrophysics Science Archive Research Center
(https://heasarc.gsfc.nasa.gov) and the \swift\ BAT 157-Month Hard X-ray
Survey page (https://swift.gsfc.nasa.gov/results/bs157mon/).
The \swift\ BAT light curve data are available at the \swift\ BAT
Hard X-ray Transient Monitor page
(https://swift.gsfc.nasa.gov/results/transients/).

%%%%%%%%%%%%%%%%%%%%%%%%%%%%%%%%%%%%%%%%%%%%%%%%%%



%\bibliographystyle{mnras}
\bibliography{tidal}

%\begin{thebibliography}{99}
%\end{thebibliography}

%%%%%%%%%%%%%%%%%%%%%%%%%%%%%%%%%%%%%%%%%%%%%%%%%%

% Don't change these lines
\bsp	% typesetting comment
\label{lastpage}
\end{document}